\documentclass[a4paper,pra,twocolumn,superscriptaddress]{revtex4-1}
\usepackage[english]{babel}
\usepackage[utf8]{inputenc}
\usepackage{graphicx,epsfig,xcolor}
\usepackage{latexsym}
\usepackage{amsfonts}
\usepackage{amsmath}
\usepackage{hyperref}
\usepackage{placeins}
\newcommand{\Tr}{\mathrm{Tr}}
\def\ket|#1>{| #1 \rangle}
\def\bra<#1|{\langle #1 |}
\def\<{\langle}
\def\>{\rangle}
\def\{{\lbrace}
\def\}{\rbrace}
\def\({\left(}
\def\){\right)}
\def\beq{\begin{equation}}
\def\eeq{\end{equation}}

\def\pl{\partial}

\def\J{\mathbf{J}}
\def\T{\mathcal{T}}
\def\ve{\varepsilon}
\def\nn{\nonumber}
\def\g{\mathbf{g}}

\begin{document}

\title{Entanglement links and the quasiparticle picture}

\author{Silvia N. Santalla}
\affiliation{Dep. de Física and Grupo Interdisciplinar de Sistemas
  Complejos (GISC), Universidad Carlos III de Madrid, Spain}

\author{Giovanni Ramírez}
\affiliation{Instituto de Investigación, Escuela de Ciencias Físicas y
  Matemáticas, Universidad de San Carlos de Guatemala, Guatemala, Guatemala}

\author{Sudipto Singha Roy}
\affiliation{Instituto de Física Teórica, UAM-CSIC, Universidad
  Autónoma de Madrid, Cantoblanco, Madrid, Spain}
\affiliation{Pitaevskii BEC Center, CNR-INO  and Department of Physics, University
  of Trento, Trento, Italy}
\affiliation{INFN-TIFPA, Trento Institute for Fundamental Physics and
  Applications, Trento, Italy}

\author{Germán Sierra}
\affiliation{Instituto de Física Teórica, UAM-CSIC, Universidad
  Autónoma de Madrid, Cantoblanco, Madrid, Spain}

\author{Javier Rodríguez-Laguna}
\affiliation{Dep. de Física Fundamental, UNED, Madrid, Spain}

\begin{abstract}
  The time evolution of a quantum state with short-range correlations
  after a quench to a one-dimensional critical Hamiltonian can be understood using the {\em quasi-particle picture}, which states that local
  entanglement spreads as if it was carried by quasi-particles which
  separate at a fixed speed. We extend the quasi-particle picture
  using the recently introduced link representation of entanglement,
  allowing us to apply it to initial states presenting long-range
  correlations. The entanglement links are current correlators, and
  therefore follow a wave equation on the appropriate configurational
  space which allows us to predict the time evolution of the
  entanglement entropies. Our results are checked numerically for free
  fermionic chains with different initial entanglement patterns.
\end{abstract}

\date{February 20, 2023}

\maketitle

\begin{figure}
  \includegraphics[width=8cm]{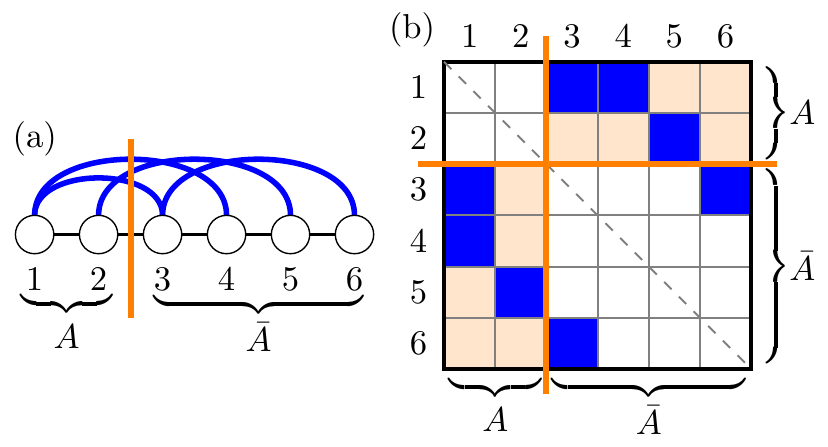}
  \caption{Illustration of the link representation. (a) Let us
    consider a quantum state containing $N=6$ units, whose non-zero
    entanglement links (EL) are denoted by the blue arches. The
    entanglement entropy (EE) between blocks $A=\{1,2\}$ and $\bar
    A=\{3,4,5,6\}$ can be obtained adding up the values of the
    non-zero EL crossing the boundary between them (orange line). (b)
    The EL can be displayed more clearly in matrix form, and the EE
    between blocks $A$ and $\bar A$ is obtained by adding up the matrix
    entries in the shaded regions above the diagonal.}
  \label{fig:illust}
\end{figure}

{\em Link representations of entanglement.-} Entanglement is a central
concept for most recent developments in quantum theory
\cite{Amico.08}. Let us consider the von Neumann entanglement entropy
(EE) $S_A=-\Tr(\rho_A \log\rho_A)$ of a block $A$, obtained from the
reduced density matrix of a pure state, $\rho_A=\Tr_{\bar
  A}|\psi\>\<\psi|$, where $\bar A$ is the complement of $A$. The
state is said to follow the area law when $S_A \propto |\partial A|$,
i.e. when the EE of a block $A$ is proportional to the measure of its
boundary \cite{Sredniki.93,Eisert.10}. The area law provides a deep connection
between entanglement and geometry \cite{Wolf.08}, paving the route for
the recent holographic approaches to a quantized gravity
\cite{Swingle.12}. Moreover, it can be generalized into the so-called
link representations of entanglement \cite{Singha.20}, which
associate to each pure state on an $N$-partite system a symmetric
matrix $J$, with $J_{ij}=J_{ji}\geq 0$, $1\leq i,j\leq N$, whose
entries are called the {\em entanglement links} (EL), such that

\beq
S_A=\sum_{i\in A,j\in \bar A} J_{ij},
\label{eq:gen_arealaw}
\eeq
for all possible blocks $A$, see Fig. \ref{fig:illust} for an
illustration. Since there are $2^N$ different blocks and the EL matrix
only contains $N(N-1)/2$ parameters, we can not ensure the existence
of an exact link representation, except in a few notable cases, such
as valence bond states \cite{Singha.21}. Yet, approximate link
representations with low errors have been found for many relevant
states. Interestingly, the entropies built from a link representation
fulfill naturally the subadditivity constraints \cite{Singha.20}.

Within a continuous framework we can write the EL as a two-point
function, $J(x,y)$, such that

\beq
S_A = \int_A dx \int_{\bar A} dy\; J(x,y).
\label{eq:SJcont}
\eeq
For a one-dimensional (1D) system, this provides an approximate route
to find the EL,

\beq
J(x,y)={1\over 2} \partial_x\partial_y S_{[x,y]},
\label{eq:Jderiv}
\eeq
where $S_{[x,y]}$ denotes the EE of the block $[x,y]$. Of course, in a
discrete setup the derivatives in Eq. \eqref{eq:Jderiv} become finite
differences,

\beq
J_{ij}={1\over 2} \( S_{i,j} - S_{i+1,j} - S_{i,j+1} + S_{i+1,j+1} \),
\label{eq:JfromS}
\eeq
where $S_{a,b}$ denotes the EE of $A=\{a,\cdots,b-1\}$. This link
representation is exact for compact blocks, losing its accuracy slowly
as the number of fragments in the partition is increased. We should
stress that the ground states (GS) of critical systems possess a very
clean link representation, with $J(x,y)\approx (c/6) |x-y|^{-2}$, thus
showing that Eq. \eqref{eq:gen_arealaw} applies to states with a
logarithmic violation of the area law
\cite{Holzhey.94,Vidal.03,Calabrese.04}, because their entanglement
entropy scales as $S_{[x,y]}\approx (c/3) \log(|x-y|/\ve)$, where
$\ve$ is the UV cutoff. This scaling suggests that the entanglement
links can be associated to the expectation value of current operators,

\beq
J(x,y) = \<\J(x)\J(y)\>,
\eeq
which is indeed the case if we let $\J(z,\bar z)=\J_L(z)+\J_R(\bar
z)$, with \cite{Singha.20}

\beq
\J_L(z)\equiv \lim_{n\to 1^+} {1\over\sqrt{2(1-n)}} \pl_z \T_n(z),
\eeq
where $\T_n(z)$ is the {\em twist field operator} of order $n$
\cite{Calabrese.04,Cardy.08,Calabrese.09}, and we define $\J_R(\bar
z)$ as the equivalent anti-holomorphic part. The scaling dimension of
the twist operators is $\Delta_n={c\over 12}(n-1/n)$, thus proving
that the entanglement current operator $\J(z)$ has scaling dimension 1
and, therefore, $\pl_{\bar z} \J_L(z)=\pl_z \J_R(\bar z)=0$. We are
therefore led to claim that

\beq
\pl_z\pl_{\bar z}\J=0,
\label{eq:zzbar}
\eeq
i.e.: the full entanglement current follows a wave equation. The
remainder of this article is devoted to exploring the consequences of
Eq. \eqref{eq:zzbar} and providing numerical checks in particular
cases. 

\begin{figure}
  \includegraphics[width=4.5cm]{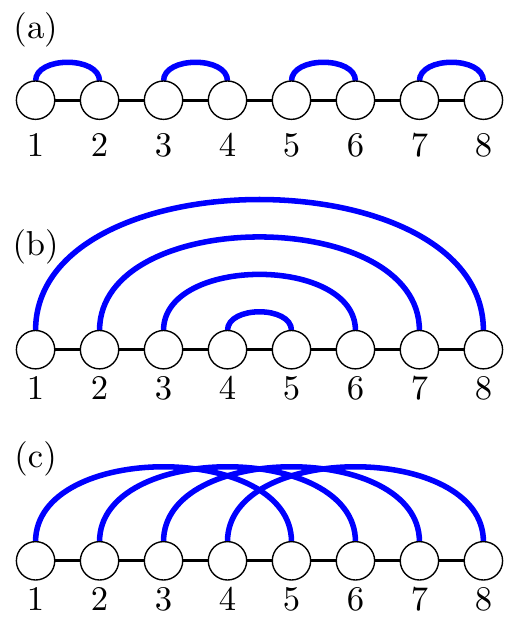}
  \caption{Illustration of the initial states considered. (a)
    Dimerized state; (b) Rainbow state; (c) Bridge state.}
  \label{fig:states}
\end{figure}

{\em Entanglement link dynamics.-} The quasi-particle picture (QPP)
successfully describes the time-evolution of the EE when a state with
short-range entanglement is quenched to a critical Hamiltonian
\cite{Calabrese.05,Calabrese.06,Calabrese.07,Fagotti.08}. In this work
we extend the quasi-particle picture using the link representation
framework, in order to characterize the time evolution of the EE of
different initial states, including those presenting long-range
entanglement. Our examples include dimerized states, see Fig.
\ref{fig:states} (a), the rainbow chain of $N$ sites
\cite{Vitagliano.10,Ramirez.14,Ramirez.15,Laguna.16,Laguna.17,Tonni.18,Samos.19,MacCormack.19,Samos.20,Samos.21},
whose GS can be approximately described as a concentric set of valence
bond states between site $i$ and site $N+1-i$, see Fig.
\ref{fig:states} (b), or the {\em bridge state} that we introduce
here, which presents bonds between sites $i$ and $i+N/2$, with $i\in
\{1,\cdots,N/2\}$, see Fig. \ref{fig:states} (c).

According to the QPP, when we quench a short-ranged state using a 1D
critical Hamiltonian the EE of a block of size $\ell=|x-y|$ at time
$t$ behaves approximately as

\beq
S(\ell,t) = \begin{cases}\sigma vt, & vt<\ell, \\
\sigma\ell, & vt>\ell,\end{cases}
\label{eq:quasip1}
\eeq
where $v$ is twice the speed of light, which may depend on the
structure of the initial state \cite{Viti.18,Singha.22}, and
$\sigma>0$ is the EE per site at saturation, and we have neglected the
initial value of the EE. This expression can be written alternatively
as

\beq
S(\ell,t)=\sigma v t\; \Theta(\ell-vt) + \sigma\ell\; \Theta(vt-\ell),
\label{eq:quasip2}
\eeq
where $\Theta(x)$ is the Heaviside function. Using
Eq. \eqref{eq:Jderiv} we can estimate the entanglement links,

\beq
J(\ell,t)=-{1\over 2}\pl^2_\ell S(\ell,t)= {\sigma\over 2} \delta(vt-\ell),
\eeq
which represents an {\em entanglement wave}. Indeed, it is
straightforward to check that this expression fulfills the wave
equation, 

\beq
{1\over v^2}\partial^2_t J(\ell,t)=\partial^2_\ell J(\ell,t),
\label{eq:waveJl}
\eeq
which can be regarded as an extension of the QPP, in accordance with
Eq. \eqref{eq:zzbar} and the requirements of conformal invariance.

Let us check the validity of Eq. \eqref{eq:waveJl} using as our
quenching Hamitonian a free-fermionic chain,

\beq
H_0=-{1\over 2}\sum_{i=1}^N c^\dagger_i c_{i+1} + \text{h.c.},
\label{eq:H0}
\eeq
with either open or periodic boundaries, as required. Notice that the
propagation velocity is $v=2$ in this case, since the Fermi velocity
is $v_F=1$. Meanwhile, our initial states will be chosen as GS of
deformed versions of that Hamiltonian

\beq
H(\g)=-\sum_{i=1}^N g_i\; c^\dagger_i c_{i+1} + \text{h.c.},
\label{eq:hamJ}
\eeq
where $\g=\{g_i\}$ are the hopping amplitudes, and we assume periodic
boundaries. GS of Hamiltonian \eqref{eq:hamJ} are Slater determinants,
whose EE can be found using single-body techniques \cite{Peschel.03}.
The EL are always numerically estimated using expression
\eqref{eq:Jderiv}, which has been shown to provide a good link
representation of these states \cite{Singha.21}.

\begin{figure}
  \includegraphics[width=8cm]{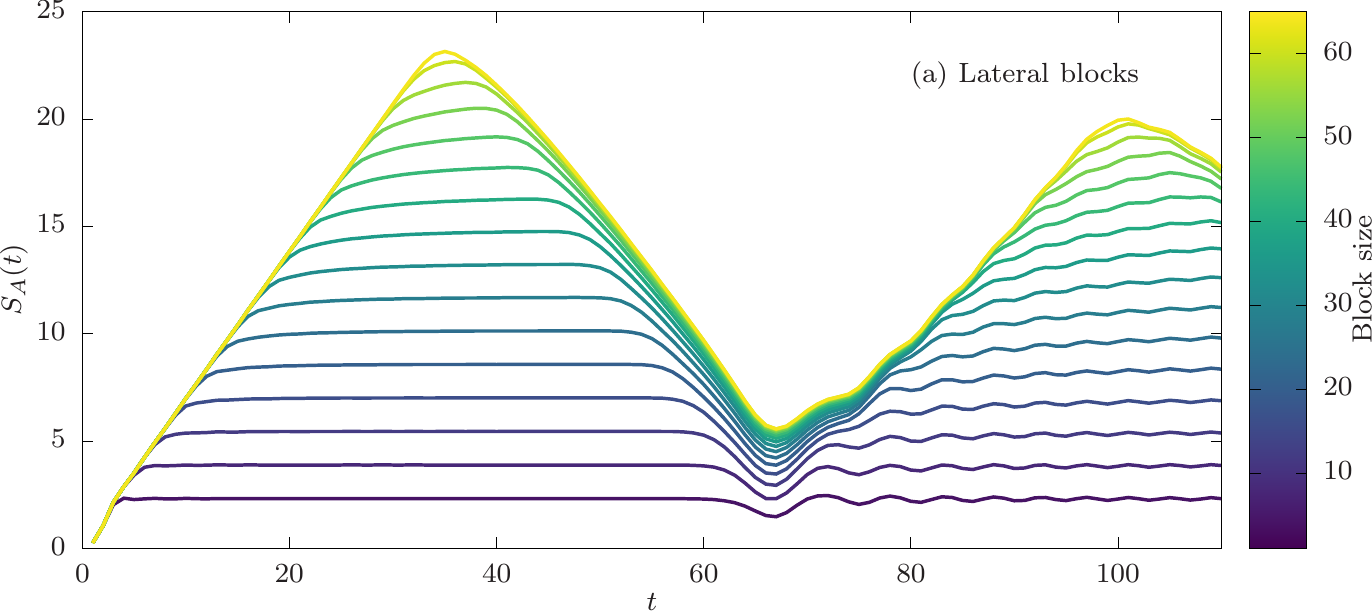}
  \includegraphics[width=4cm]{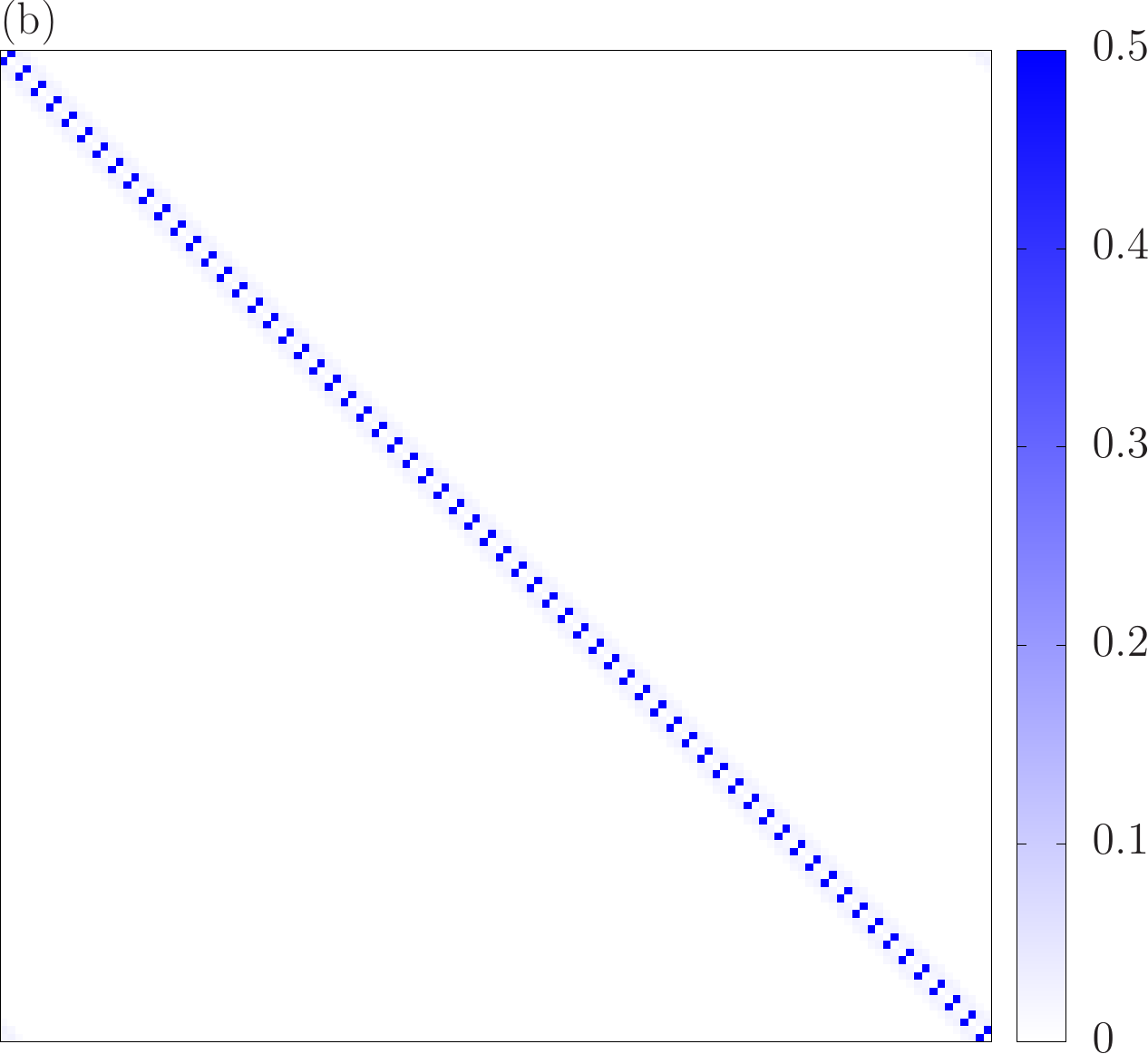}
  \includegraphics[width=4cm]{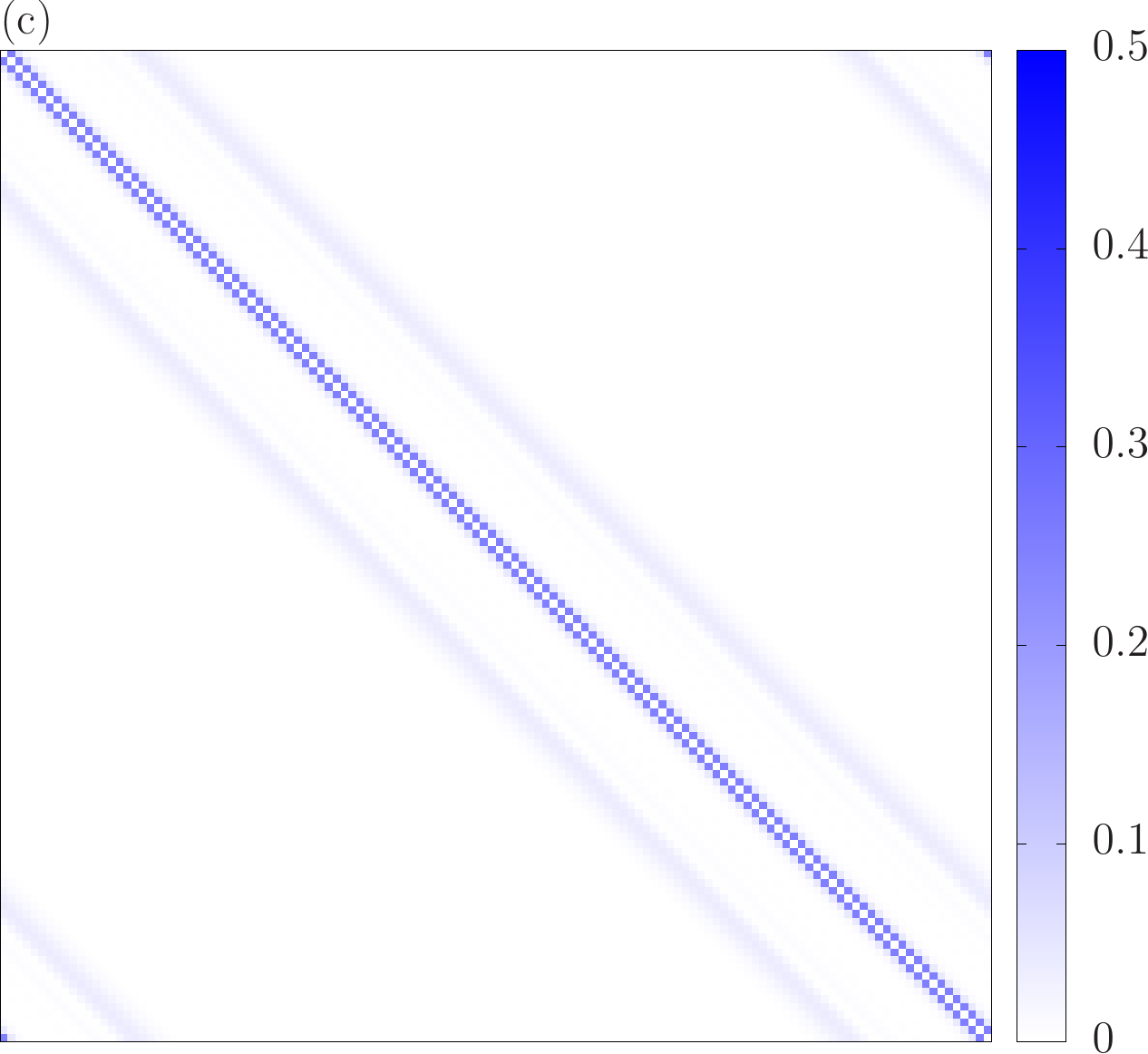}
  \caption{(a) Time evolution of the EE of the dimerized state with
    $\delta=0.5$ and $N=128$. (b) EL matrix for the dimerized state at
    time $t=0$. (c) EL matrix, $J(x,y)$, for the dimerized state
    quenched to the homogeneous Hamiltonian, after a time
    $t=10$. Notice the two wavefronts above the main diagonal,
    traveling in opposite directions, due to the periodic boundary
    conditions.}
  \label{fig:dimer}
\end{figure}

Let us consider a dimerized initial state, defined by
$g_i=(1+(-1)^i\delta)$, using $\delta=1/2$ and $N=128$. Fig.
\ref{fig:dimer} (a) shows the time evolution of the EE of blocks of
different sizes $\ell$. We observe that each block saturates at a time
proportional to its size, as proposed by the QPP. In Fig.
\ref{fig:dimer} (b) we can see the EL matrix of the initial state. The
only non-zero EL are located near the diagonal, $J_{i,i+1}=J_{i+1,i}$.
Fig. \ref{fig:dimer} (c), on the other hand, provides the EL matrix at
time $t=10$, showing that two wavefronts are traveling in opposite
directions from the original diagonal, as proposed in Eq.
\eqref{eq:waveJl} if the initial velocity, $\pl_t J(\ell,t=0)$, is
zero. 

\begin{figure}
  \includegraphics[width=8cm]{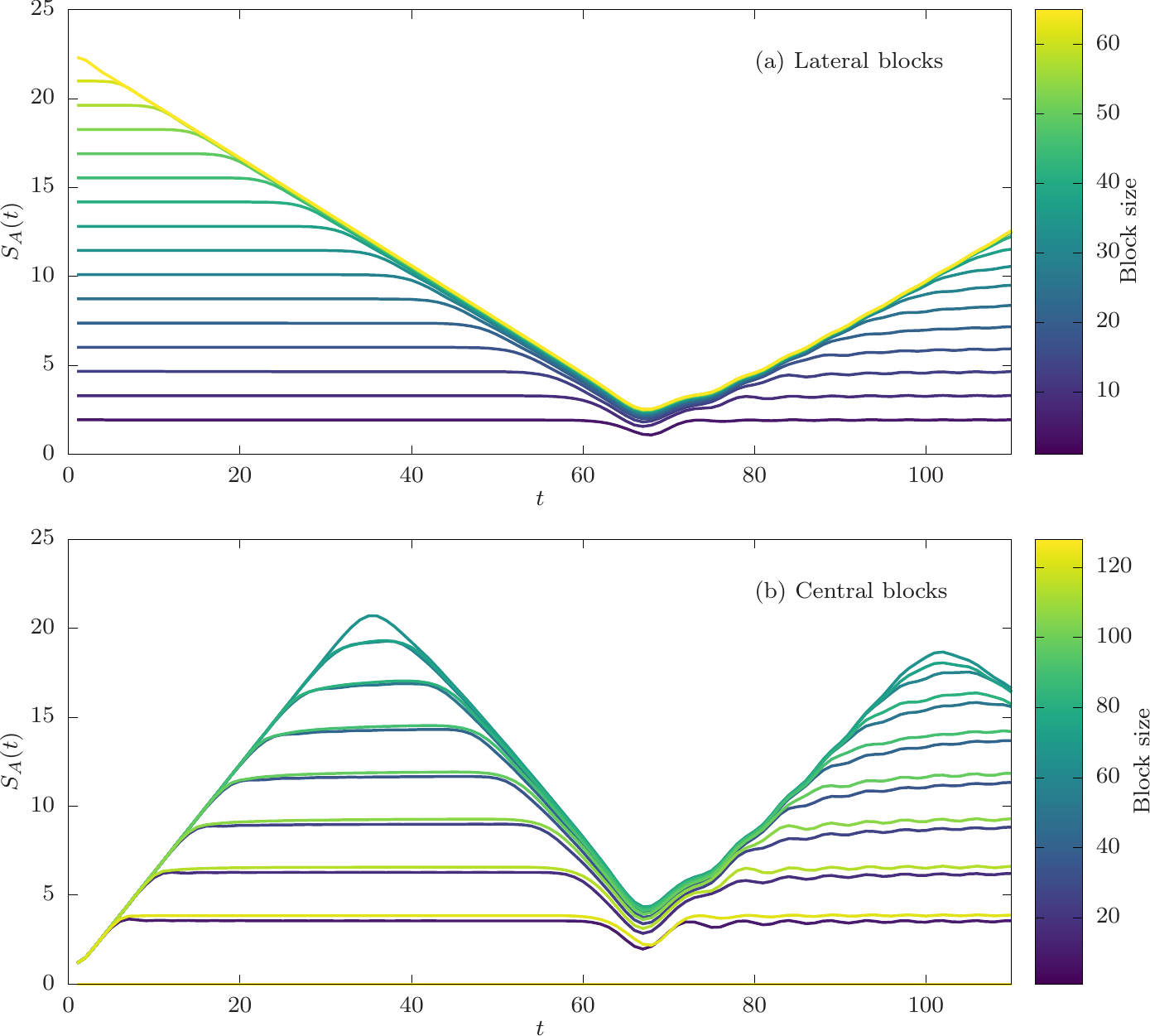}
  \includegraphics[width=4cm]{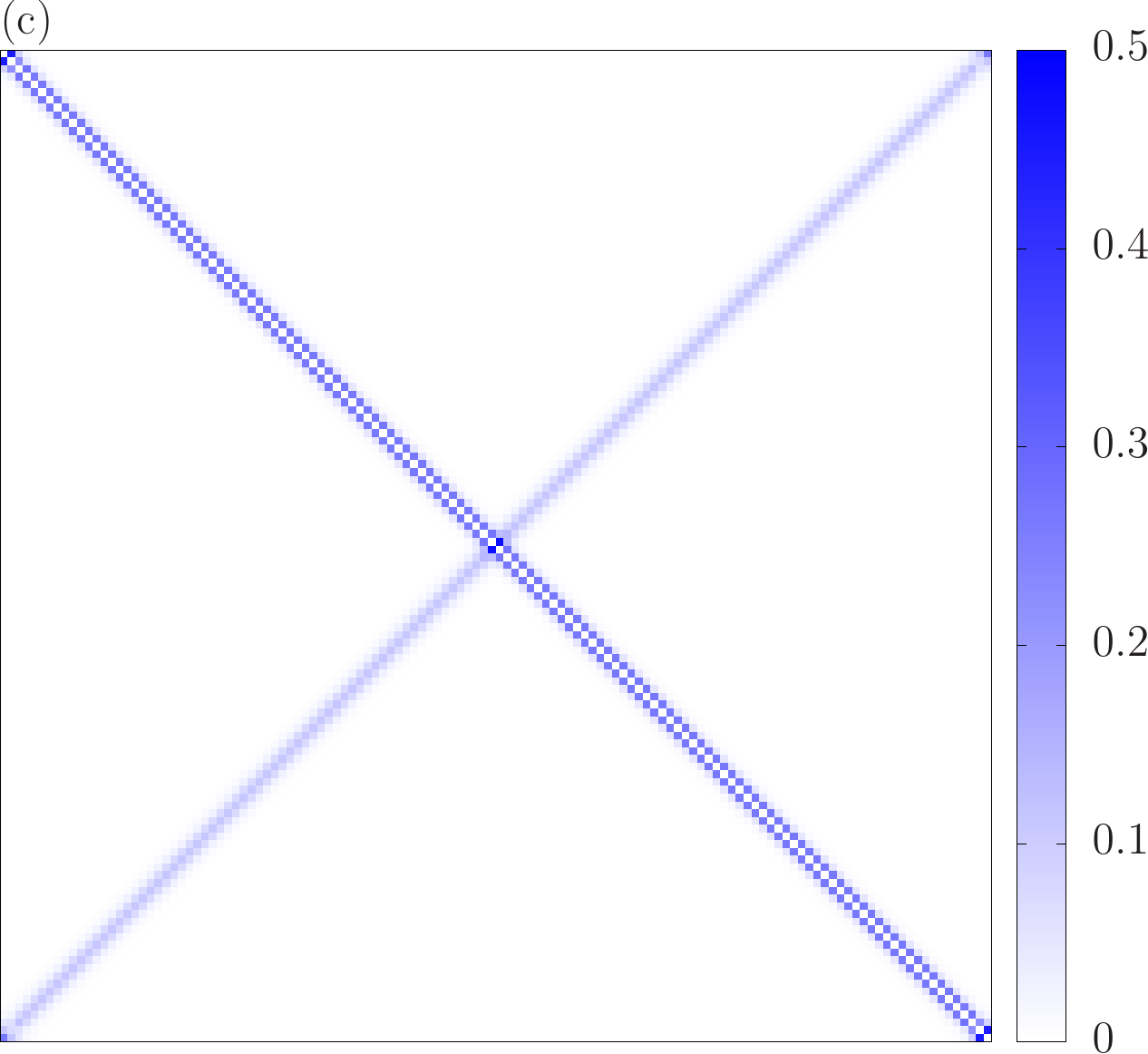}
  \includegraphics[width=4cm]{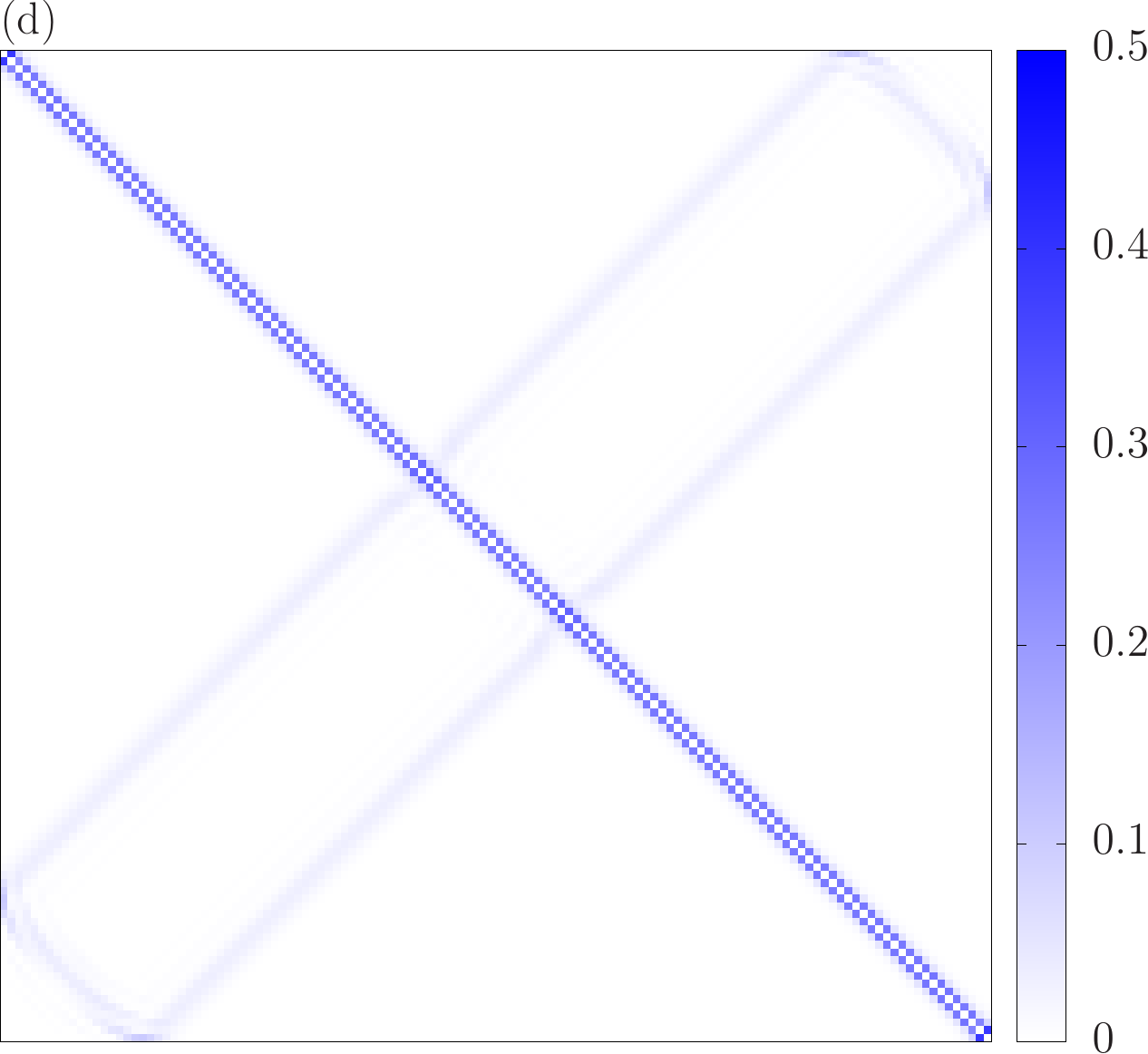}
  \caption{(a) Time evolution of the EE of lateral blocks within the
    rainbow state, (b) same figure for central blocks. (c) EL matrix
    for the rainbow state at $t=0$, using $h=0.7$ and $N=128$, (d) EL
    matrix after $t=10$, under a quench to an open boundaries
    homogeneous Hamiltonian $H_0$.}
  \label{fig:rainbow}
\end{figure}

{\em EE of the rainbow state.-} Let us apply the QPP to states out of
its initial range of applicability, such as the GS of the free-fermionic
rainbow chain
\cite{Vitagliano.10,Ramirez.14,Ramirez.15,Laguna.16,Laguna.17,Tonni.18,Samos.19,MacCormack.19,Samos.20,Samos.21},
which presents long-range entanglement and is defined by Hamiltonian
\eqref{eq:hamJ} using

\beq
g_i = \begin{cases} 1 & \text{if } i=N/2,\\
  e^{-h(|N/2-i|-1/2)} & \text{otherwise.}
\end{cases}
\eeq
The EE of blocks within the GS of the rainbow Hamiltonian grows
linearly with their size, $S(\ell) \approx h\ell/6$ for $h\gg 1$, and
site $i$ is most strongly correlated to site $N+1-i$, thus showing a
concentric bond structure which justifies the rainbow term, see Fig.
\ref{fig:states} (b). We will take as our example the case $N=128$ and
$h=0.7$, which is quenched to the homogeneous Hamiltonian $H_0$,
although in this case we will choose open boundaries. The time
evolution of the EE of the lateral blocks $A_\ell=[1,\cdots,\ell]$ is
shown in Fig. \ref{fig:rainbow} (a). At time $t=0$ the entanglement
entropy of these blocks is proportional to their length, as the volume
law requires. The EE of the largest one, $\ell=N/2$, starts to
decrease immediately after the quench. Yet, the EE of smaller lateral
blocks remains constant for a certain time that increases linearly as
the block size decreases. Interestingly, the block of size $\ell$
starts its decrease when the entropy of all larger blocks reach its
starting value. After all the blocks have reached their minimum value
they start growing linearly, in similarity to the dimerized state,
saturating at a value proportional to their size, corresponding
approximately to their initial value.

We may also consider the EE of central blocks of size $\ell$, whose
initial entanglement is approximately zero, as we can see in
Fig. \ref{fig:rainbow} (b). They start growing immediately, twice as
fast as the lateral blocks decrease.

{\em Extended predictions of the QPP.-} It is relevant to ask whether
the quasi-particle picture can provide an explanation for these
observations. Indeed, based on Eq. \eqref{eq:zzbar}, we propose the
following wave equation

\beq
    {1\over v^2} \partial^2_t J(x,y) =
    {1\over 2} \(\partial^2_x+\partial^2_y\)  J(x,y),
\label{eq:wave_j}
\eeq
with suitable boundary conditions, in order to explain the time
evolution of both short-ranged and long-ranged initial states. Notice
that, due to the symmetry $J(x,y)=J(y,x)$, we could have used only
$\pl^2_x$ or $\pl^2_y$. Let us apply this equation to the rainbow
case, with open boundaries. In Fig. \ref{fig:rainbow} (c) we can see
the EL matrix for $t=0$, which contains two different features: a
diagonal double line, corresponding to the local entanglement, and an
opposite diagonal line, which shows the links between sites $i$ and
$N+1-i$, which conforms the (approximate) concentric bonds. The time
evolution leaves the first line invariant, but it splits the second
one into two wavefronts, one of which propagates rightwards and the
other one leftwards, as we can see in the EL matrix for $t=10$ in Fig.
\ref{fig:rainbow} (d). Moreover, we see a new wavefront appear,
parallel to the main diagonal, joining the previous two fronts.

We note that the long-range EL shown in Fig. \ref{fig:rainbow} (c) and
(d) are solutions of the entanglement wave equation,
Eq. \eqref{eq:wave_j}, if we choose as initial conditions
$J(x,y,0)=\sigma\delta(x+y-N)$ and $\pl_t J(x,y,0)=0$, leading to

\begin{align}
  J(x,y,t) =&{\sigma\over 2} \delta(x+y-N+vt)
  + {\sigma\over 2}\delta(x+y-N-vt) \nn\\
  +& {\sigma\over 2} \delta(x-y-N+vt),
  \label{eq:Jrainbow}
\end{align}
where the last term corresponds to a line parallel to the main
diagonal joining the two other lines, due to reflection at the
boundaries. Indeed, this is what we can observe to a good
accuracy. Yet, the main diagonal of the EL matrix remains static,
showing a remanent non-universal behavior.

How can we understand Eq. \eqref{eq:Jrainbow} in physical terms? Using
Huygens principle, we start out with the rainbow wavefront, which
corresponds to the secondary diagonal, $y=N-x$. Each site of the
wavefront generates a new circular wave around it, and the new
wavefront is given by their envelope. In other terms: each EL {\em
  excites} the neighboring links: one slightly stretched, one slightly
shrunk and two slightly displaced leftwards and rightwards. Yet, for
the rainbow system the stretched and shrunk links are already excited,
so only the translated links appear. Of course, the boundary
conditions should also be taken into account: the largest link can not
merely translate, so it gives rise to two links, one in which the left
extreme bounces rightwards and another one in which the right extreme
bounces leftwards.

\begin{figure}
\includegraphics[width=8.5cm]{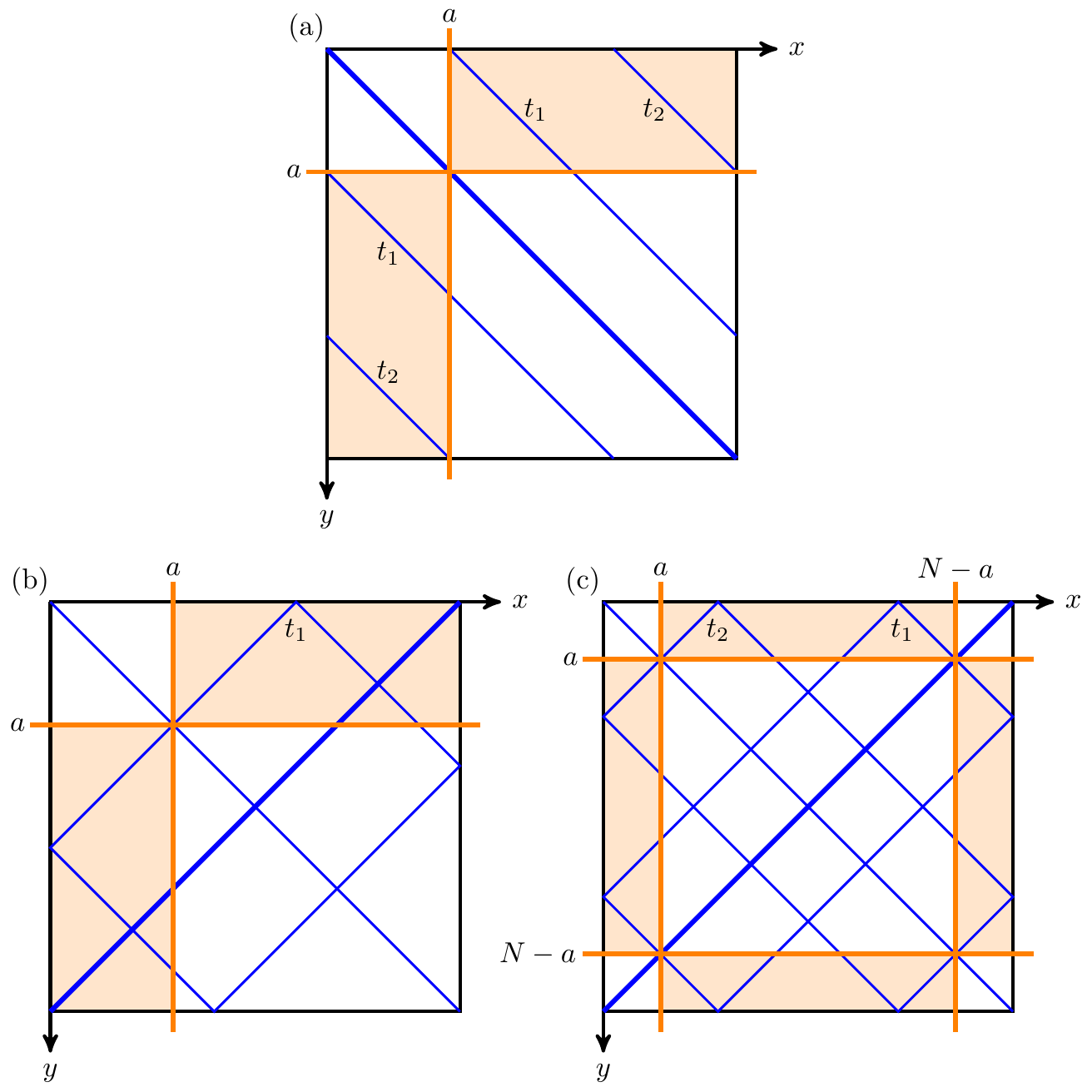}
\caption{Illustrating the evaluation of the EE for any time, given the
  time evolution of the EL. The strong blue line denotes the EL at
  time $t=0$, and the parallel thin lines correspond to the EL at any
  later times, traveling with constant speed. (a) Dimerized case. The
  EE of the block $A=[0,a]$ can be evaluated by counting the EL in the
  shaded area. Notice that, due to the periodic boundaries, the front
  is formed by two apparently disjoint segments. (b) Rainbow case,
  using the same block $A=[0,a]$. (c) Rainbow case, using lateral
  blocks, $B=[a,N-a]$.}
\label{fig:graphical}
\end{figure}

Integration of Eq. \eqref{eq:Jrainbow} according to
Eq. \eqref{eq:SJcont} will provide the time-evolution of the EE of any
block \cite{Calabrese05_Eq41}. Yet, we would like to show a graphical
procedure that will yield a better physical intuition in
Fig. \ref{fig:graphical}. For the sake of clarity, let us start with
an initial state with short-range entanglement, in
Fig. \ref{fig:graphical} (a), where we have neglected boundary effects
for simplicity. The initial entanglement is denoted by the thick blue
line, that splits into two lines traveling in opposite directions. In
general terms, the EE obtained when a diagonal line crosses a
rectangle is given by the projection of that line on any of the
axes. Thus, we see that the EE grows linearly up to time $vt_1=a$,
remains constant up to $vt_2=N-a$, and decays linearly down to zero
from that moment on \cite{PBC}. That is the prediction of the QPP,
given in Eq. \eqref{eq:quasip1}.

Let us consider the EL of the rainbow state with open boundaries,
schematically drawn in Fig. \ref{fig:graphical} (b) for lateral blocks
$A=[0,a]$ with $a<N/2$. Notice that the EE will remain constant up to
time $vt_1=N-2a$. Then it will start decreasing linearly until it
reaches zero at time $vt=N$. In other terms, we get an expression
similar to Eq. \eqref{eq:quasip1},

\beq
S_A(t)=\begin{cases}
\sigma a, & vt<vt_1=N-2a, \\
\sigma (a - v(t-t_1)/2), & vt_1<vt<N.
\end{cases}
\eeq
Notice that if $a>N/2$ we get the same EE as with the block with size
$N-a$. Let us now do the same calculation for central blocks of the
form $B=[a,N-a]$, with $a<N/4$, using Fig. \ref{fig:graphical} (c).
Notice that the EE starts at zero, grows linearly until time
$vt_1=a$. Then it remains constant until time $vt_2=N-a$, when it will
start a linear decrease to zero at time $vt=N$. In other terms,

\beq
S_B(t)=\begin{cases}
\sigma vt/2, & vt<vt_1=2a,\\
\sigma a, & vt_1<vt<vt_2=N-2a,\\
\sigma (a-v(t-t_2)/2), & vt_2<vt<N.
\end{cases}
\eeq
If $a>N/4$, our graphical procedure shows that the time evolution of
the EE of the block $[a,N-a]$ corresponds to that of block
$[N/2-a,N/2+a]$, which is approximately the case as we can see in
Fig. \ref{fig:rainbow} (b).

\begin{figure}
  \includegraphics[width=8cm]{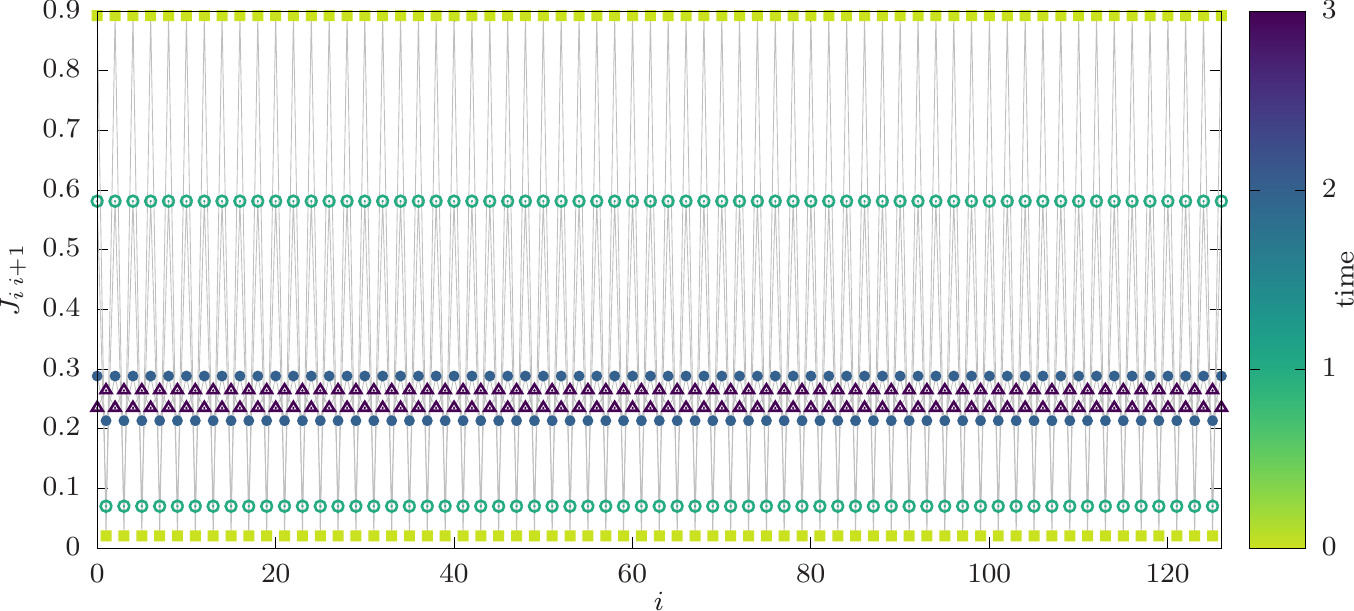}
  \includegraphics[width=8cm]{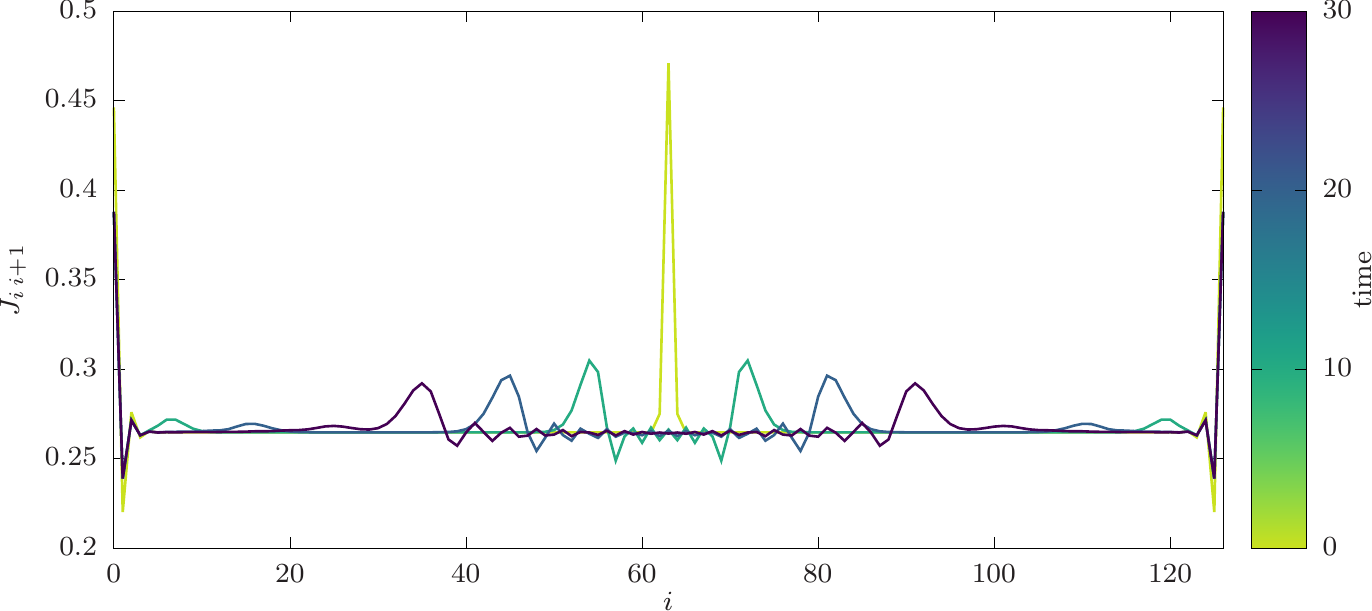}
  \caption{Sub-diagonal EL of the dimerized and rainbow cases shown in
    Fig. \ref{fig:dimer} and \ref{fig:rainbow}, as a function of time.}
  \label{fig:subdiagonal}
\end{figure}

{\em The persistent sub-diagonal.-} Figures \ref{fig:dimer} (c) and
\ref{fig:rainbow} (d) show a persistent sub-diagonal line in the EL
matrix for longer times, which seems to challenge our description.
Yet, we should notice that Eq. \eqref{eq:wave_j} should be
complemented with suitable boundary conditions and probably a source
term along the $x=y$ line, which is privileged by the local structure
of the quenching Hamiltonian, $H_0$. Thus, we conjecture that the
persistent sub-diagonal is a non-universal phenomenon. Indeed, Fig.
\ref{fig:subdiagonal} shows the behavior of links $J_{i,i+1}$ for the
dimerized and rainbow cases analyzed before as a function of time,
thus linked to Fig. \ref{fig:dimer} and \ref{fig:rainbow}. In the
dimerized case, Fig. \ref{fig:subdiagonal} (a), the sub-diagonal EL
tend to a constant after a quick transient. In the rainbow case, Fig.
\ref{fig:subdiagonal} (b), we observe a wave pulse propagating over a
constant background.

\begin{figure}
  \includegraphics[width=8cm]{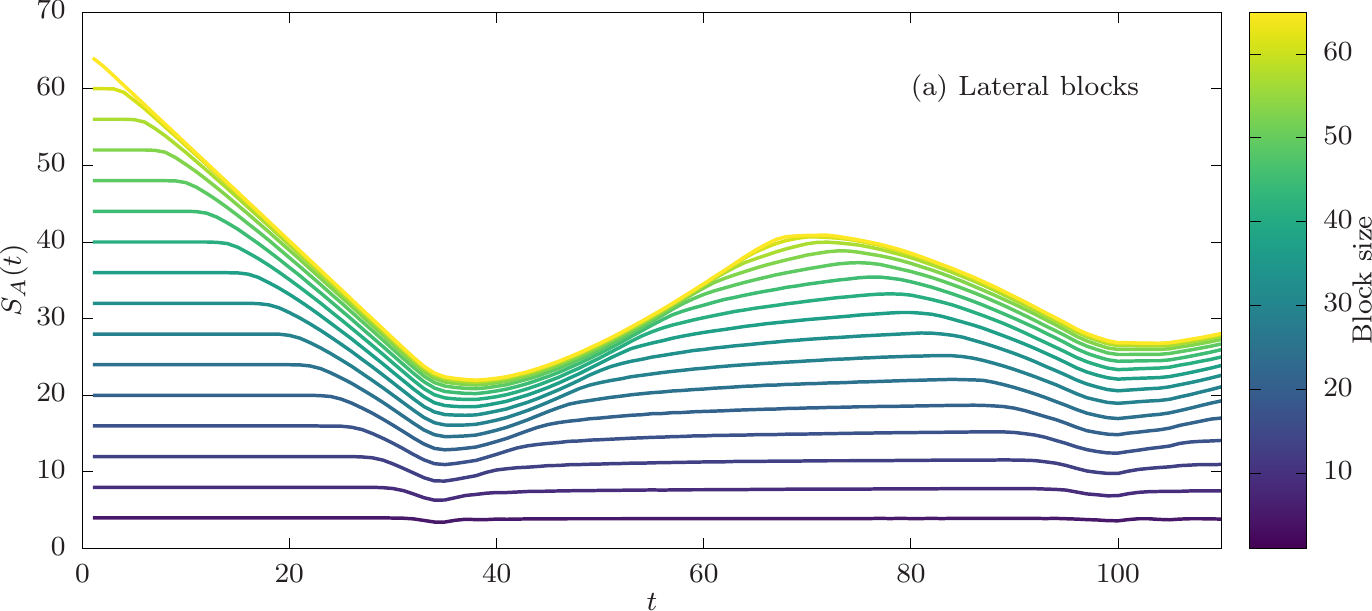}
  \includegraphics[width=4cm]{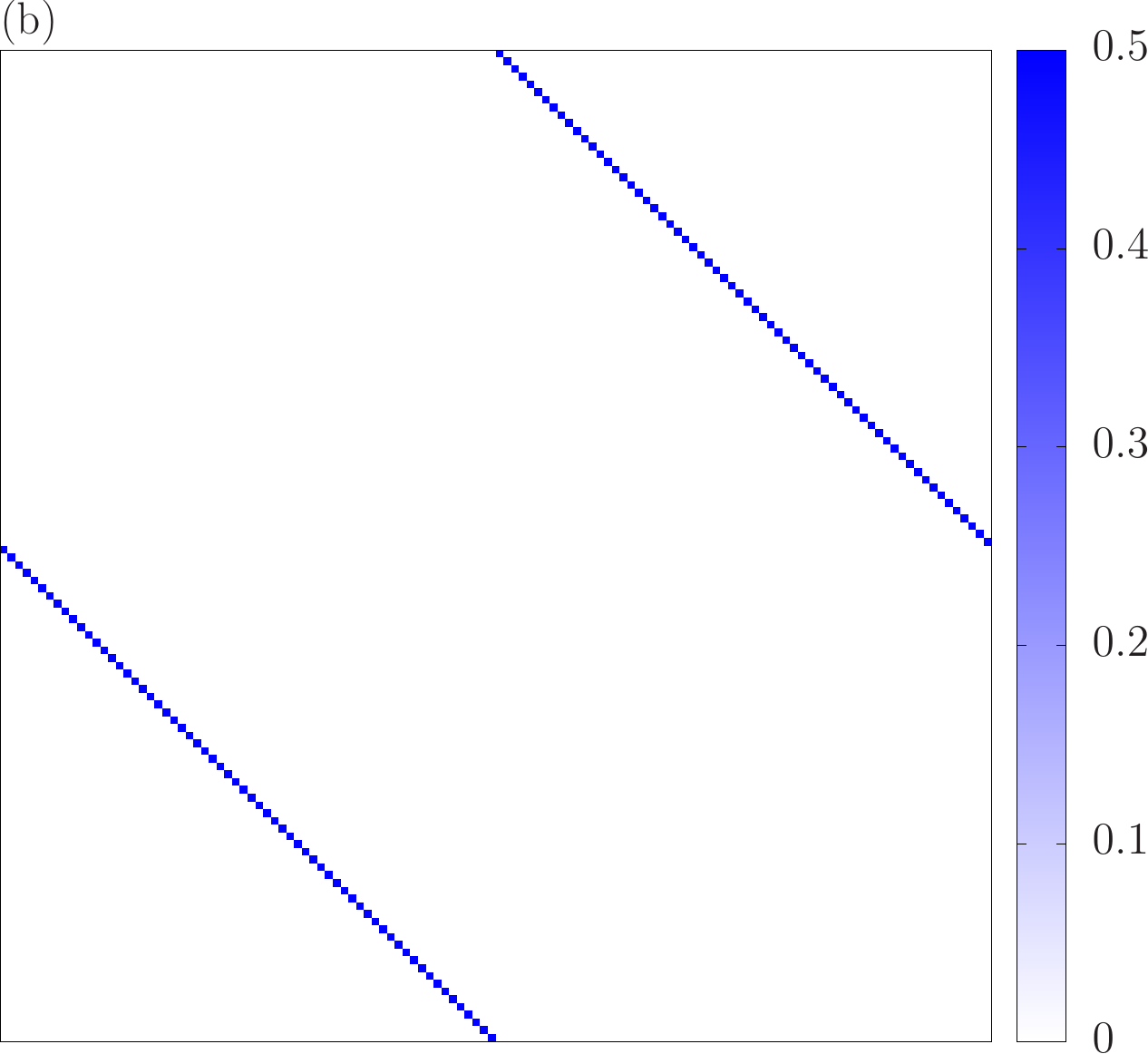}
  \includegraphics[width=4cm]{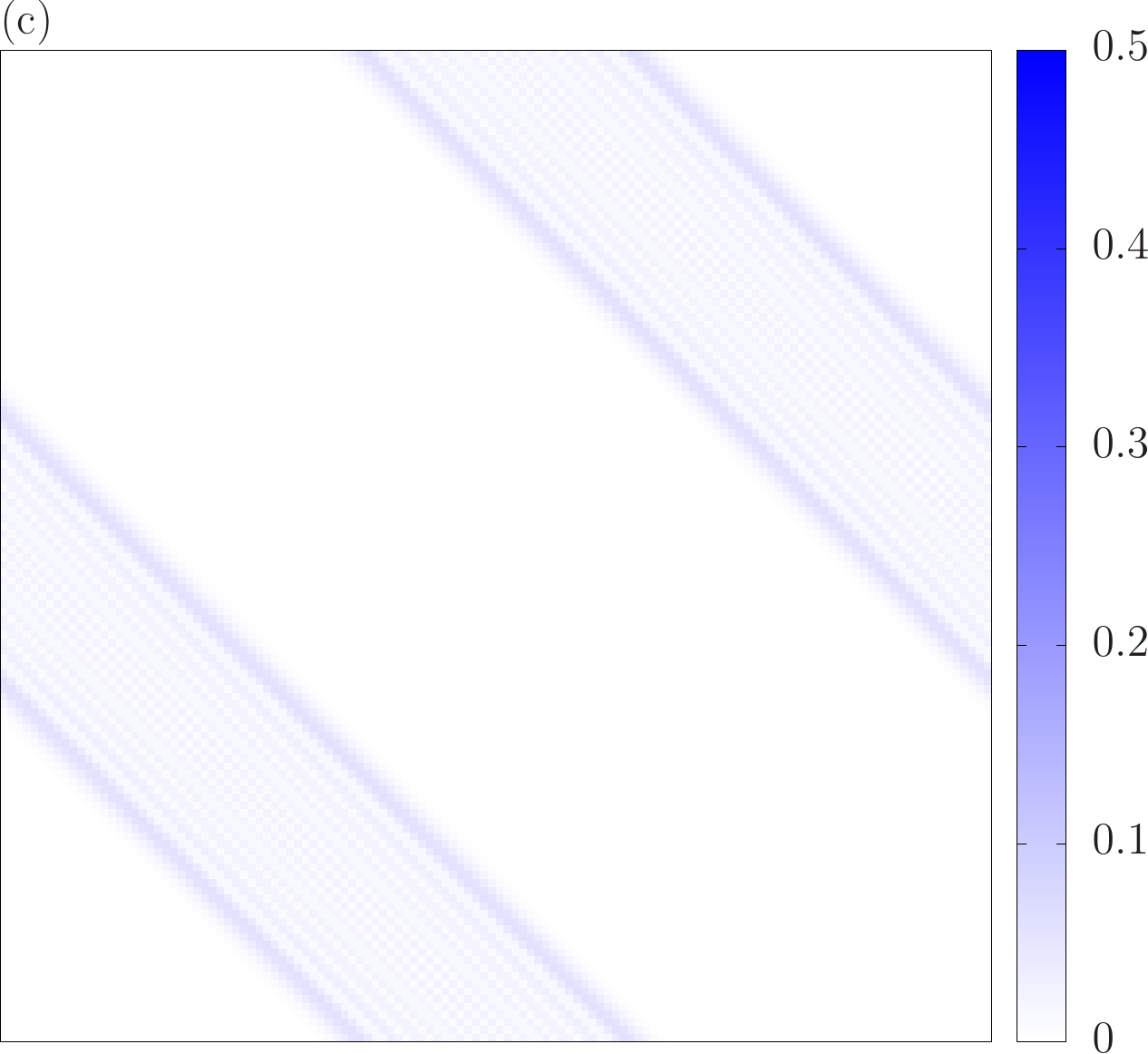}
  \caption{(a) Time evolution of the EE of contiguous blocks within
    the bridge state of $N=128$ sites. Since the state and the
    quenching Hamiltonian are translation invariant, all contiguous
    blocks of the same size have the same entropy. (b) EL matrix at
    time $t=0$. (c) EL matrix after $t=10$.}
  \label{fig:bridge}
\end{figure}

{\em The bridge state.-} Let us consider one last example, a valence
bond state that we will call the {\em bridge state},

\beq
\ket|\Psi>= \prod_{k=1}^{N/2} {1\over\sqrt{2}}\(c^\dagger_k + (-1)^k
c^\dagger_{k+N/2}\) \ket|0>,
\label{eq:bridge}
\eeq
where $\ket|0>$ is the Fock vacuum, see Fig. \ref{fig:states} (c).
Now, let us show that the evolution of state $\ket|\Psi>$ after a
quench to $H_0$ (with periodic boundaries) can be described using the
extended version of the QPP. At $t=0$, the state is a valence bond
solid, and thefore the EL matrix can be exactly found, $J_{ij}=\sigma
\delta_{i,j+N/2 \mod N}$ \cite{Singha.21}. We may solve the wave
equation with initial condition $J(x,y)=\sigma \delta(x-y+N/2)$ and
$\pl_t J(x,y)=0$, and obtain two traveling waves, $J(x,y,t)=(\sigma/2)
\delta(x-y+N/2+vt) + (\sigma/2) \delta(x-y+N/2-vt)$, as we can observe
in Fig. \ref{fig:bridge} (a) and (b). Yet, in this case the wavefronts
leave a larger amount of {\em radiation} behind, since the lattice
artifacts are stronger because the initial state is synthetic. We are
then led to predict that the entanglement entropy behaves as

\beq
S(\ell,t)=\begin{cases}\sigma \ell, & vt<N/2-\ell, \\
\sigma(N/2-vt), & N/2-\ell<vt<N/2,\end{cases}
\eeq
before the revivals due to the periodic boundaries, which corresponds
to the results shown in Fig. \ref{fig:bridge}.

{\em Conclusions and further work.-} We have extended the
quasi-particle picture (QPP) to describe the time evolution of the
entanglement entropy after a quench to a critical Hamiltonian, when
the initial state presents long-range entanglement. Our extended
description of the QPP makes use of the entanglement link (EL)
representation for the entanglement entropy of different blocks.
Conformal field theory arguments show that the EL must fulfill a wave
equation on a tensor-product space built by two copies of the
configuration space, due to their nature as current correlators. The
propagation velocity associated to the wave equation depends both on
the quenching Hamiltonian and the structure of the initial state
\cite{Viti.18,Singha.22}, and may even vanish, e.g. in the case of
eigenstates of the quenching Hamiltonian. We should stress that our
predictions only apply to 1+1D critical Hamitonians, for any value of
their central charge, but they need not apply to non-critical cases,
in which another evolution equation should be expected. Moreover, we
have observed non-universal effects near the diagonal line of the EL
matrix, which are in need of further clarification.

Our results have been checked for the case of free fermions on a
chain, using initial states with different patterns of entanglement.
More numerical experiments are required to check the validity of our
formulation for other conformally invariant systems, such as the
critical Ising or XXZ models. Moreover, our predictions refer to the
EL obtained from the von Neumann entropy. Higher-order Rényi entropies
give rise to alternative EL representations which may present a
different time-evolution and are also interesting to investigate.

Of course, this extension of the QPP is subject to lattice effects
which limit its validity in the long run \cite{Prahofer.02,Singha.22}.
It is relevant to ask how these lattice effects will spoil its
predictions for different initial states and different critical
models. It is also interesting to wonder about the application of our
extension of the QPP to higher dimensional systems, where we can find
phenomena such as the {\em entanglement tsunami} that describes the
evolution of the EE in certain holographic setups \cite{Liu.14}.


\begin{acknowledgments}
  We would like to thank P. Calabrese and E. Tonni for very useful
  discussions. We acknowledge the Spanish government for financial
  support through grants PGC2018-095862-B-C21, PGC2018-094763-B-I00,
  PID2019-105182GB-I00, PID2021-123969NB-I00, PID2021-127726NB-I00,
  QUITEMAD+ S2013/ICE-2801, SEV-2016-0597 of the ``Centro de
  Excelencia Severo Ochoa'' Programme and the CSIC Research Platform
  on Quantum Technologies PTI-001. We also acknowledge support by the
  ERC Starting Grant StrEnQTh (project ID 804305), Provincia Autonoma
  di Trento, and by Q@TN, the joint lab between University of Trento,
  FBK-Fondazione Bruno Kessler, INFN-National Institute for Nuclear
  Physics and CNR-National Research Council.
\end{acknowledgments}


\begin{thebibliography}{999}
\bibitem{Calabrese05_Eq41} We would like to stress the similarity
  between Eq. (4.1) of \cite{Calabrese.05} and the integration of
  Eq. \eqref{eq:Jrainbow} according to Eq. \eqref{eq:SJcont}. Indeed,
  that equation provides a link representation of the time-evolved
  state within the QPP.
\bibitem{PBC} Notice that, due to the periodic boundaries, the
  wavefront will always be formed by two lines parallel to the main
  diagonal traveling with opposite velocities. In fact, the two lines
  in Fig. \ref{fig:graphical} (a) corresponding to times $t_1$ and
  $t_2$ are present at both times, interchanging their velocities in
  the process. Check Fig. \ref{fig:dimer} (c) where we can see both
  wavefronts at an early time.
\end{thebibliography}
\end{document}